# Electrically-gated near-field radiative thermal transistor

Yue Yang and Liping Wang*

*School for Engineering of Matter, Transport, and Energy*
*Arizona State University, Tempe, AZ, 85287 USA*

In this work, we propose a near-field radiative thermal transistor made of two graphene-covered silicon carbide (SiC) plates separated by a nanometer vacuum gap. Thick SiC plates serve as the thermal "source" and "drain", while graphene sheets function as the "gate" to modulate the near-field photon tunneling by tuning chemical potential with applied voltage biases symmetrically or asymmetrically. The radiative heat flux calculated from fluctuational electrodynamics significantly varies with graphene chemical potentials, which can tune the coupling between graphene plasmon across the vacuum gap. Thermal modulation, switching, and amplification, which are the key features required for a thermal transistor, are theoretically realized and analyzed. This work will pave the way to active thermal management, thermal circuits, and thermal computing.

*Electronic mail: liping.wang@asu.edu

In analogy to the usual electrical transistor for the control of electric current, thermal transistor, which consists of thermal source, gate and drain components, is used to control heat flow. The first model of thermal transistor was proposed by Li through phonon transport [1, 2]. Compared to phonon, photon transport has its advantage of much faster speed. A thermal transistor based on near-field radiation was also proposed by placing the phase transition vanadium dioxide ($VO_2$) thin film between two $SiO_2$ plates [3]. However, this three-body thermal transistor system is extremely challenging to achieve in the near-field experimentally, especially for the suspended gate with nanometer vacuum gaps from the source and the drain. The phase transition of $VO_2$ is also limited to the temperature range of only several degrees, and in order to achieve the thermal transistor function, the temperature of $VO_2$ gate must be maintained at the specific phase transition range. In addition, near-field thermal rectifiers and switch [4-8] have been constructed as well for thermal management.

In this study, a near-field radiative thermal transistor made of two graphene-covered SiC plates separated by a nanometer vacuum gap is proposed. Thick SiC plates serve as the thermal "source" and "drain", while graphene sheets function as the "gate". As a gapless two-dimensional (2D) semiconductor, graphene supports surface plasmon in the terahertz and infrared region and its optical property can be tuned by changing the chemical potential through applying external voltage biases [9-12]. People have already made attempts to apply graphene to tune the near-field radiative heat transfer [13-16]. Recently, graphene has also been used to enhance the near-field radiative heat transfer between silica gratings [17].

Compared to the VO2 based near-field thermal transistor, without the limit for gate temperature, the graphene based one has more flexibility. Instead of thermally changing the gate temperature, it should be faster and more convenient to change graphene chemical potential electrically through applied voltage biases. From the experiment aspect, it is much easier to put two parallel plates together with nanometer vacuum gap rather than three plates. Recent progress has been made on near-field radiation measurement between two plates separated by nanometer gaps [18-20].

As the configuration shows in Fig. 1, the SiC plate with higher temperature ($T_S$ = 400 K) and that with low temperature ($T_D$ = 300 K) represent the thermal "Source" and "Drain", respectively. The gap distance between them is set as $d$ = 10 nm. Different from the traditional thermal transistor, which has one separated thermal gate, the two graphene sheets function as the electrical gate in this study. Instead of controlling thermal transistor through the gate temperature change, we can achieve it through varying chemical potentials of these two graphene sheets, which can be tuned via external voltage bias ($V_{GS}$ for source graphene and $V_{GD}$ for drain graphene) presented in Fig. 1. Two metal plates covering on the top of thermal source and the bottom of thermal drain are used as the ground electrodes.

Fluctuational electrodynamics [21], based on the stochastic nature of thermal emission, is used to calculate the near-field radiative heat fluxes. The analytical expression for the spectral radiative heat flux between two semi-infinite homogeneous media at temperatures of $T_S$ and $T_D$ is [22, 23].

$$q''_\omega = \frac{1}{4\pi^2}\left[\Theta(\omega,T_S) - \Theta(\omega,T_D)\right]\int_{\omega/c}^{\infty} \xi^p_{\text{evan}}(\omega,\beta)\beta d\beta \qquad (1)$$

where $\Theta(\omega,T) = \hbar\omega/[\exp(\hbar\omega/k_B T) - 1]$ is the mean energy of a Planck oscillator. Only the contribution from p polarized evanescent waves, which dominate the heat transfer at small vacuum gaps when surface plasmon polaritons are present, are considered here. The transmission probability function $\xi^p_{\text{evan}}$ can be written as [22, 23]

$$\xi_{\text{evan}}^{p}(\omega,\beta) = \frac{4\,\text{Im}(r_{01}^{p})\,\text{Im}(r_{02}^{p})\,e^{-2\text{Im}(\gamma_{0})d}}{\left|1 - r_{01}^{p}r_{02}^{p}e^{i2\gamma_{0}d}\right|^{2}} \quad (2)$$

where $c$ is the light velocity in vacuum, $\beta$ is the component of wavevector parallel to the interface, $\gamma_{i} = \sqrt{\varepsilon_{i}\omega^{2}/c^{2} - \beta^{2}}$ is the component of wavevector vertical to the interface in medium $i$, $\varepsilon_{i}$ is the relative dielectric function of medium $i$, $r_{0i}$ is the Fresnel reflection coefficient from vacuum to medium $i$. The subscripts 0, 1, 2 represent the vacuum and two different semi-infinite plates, respectively. As shown in Fig. 1, when the substrates are covered with graphene sheets, the Fresnel reflection coefficient at the interface between vacuum and medium $i$ separated by a monolayer of graphene can be expressed as follows with the graphene surface conductivity [10, 11, 15].

$$r_{ij}^{p} = \frac{\varepsilon_{i}\gamma_{0} - \gamma_{i} + \sigma(\omega)\gamma_{0}\gamma_{i}/(\omega\varepsilon_{0})}{\varepsilon_{i}\gamma_{0} + \gamma_{i} + \sigma(\omega)\gamma_{0}\gamma_{i}/(\omega\varepsilon_{0})} \quad (3)$$

where $\varepsilon_{0}$ and $\mu_{0}$ are the absolute electric permittivity and magnetic permeability of vacuum, respectively. $\sigma$ is the electrical conductivity of graphene sheet [9, 10, 15, 24]. The SiC dielectric function is obtained from Lorentz model [22].

$$\varepsilon_{\text{SiC}}(\omega) = \varepsilon_{\infty}\left(1 + \frac{\omega_{\text{LO}}^{2} - \omega_{\text{TO}}^{2}}{\omega_{\text{TO}}^{2} - i\gamma\omega - \omega^{2}}\right) \quad (4)$$

where $\omega$ is the frequency in wavenumber, the high-frequency constant $\varepsilon_{\infty} = 6.7$, the longitudinal optical-phonon frequency $\omega_{\text{LO}} = 969$ cm$^{-1}$, the transverse optical-phonon frequency $\omega_{\text{TO}} = 793$ cm$^{-1}$, and scattering rate $\gamma = 4.76$ cm$^{-1}$.

Spectral heat fluxes between thermal source and drain are shown in Fig. 2 with the same chemical potentials applied to both graphene sheets, which is denoted as the symmetric case.

Two peaks with magnitude above $q_\omega = 10$ nJ/m$^2$-rad can be clearly observed on the spectral heat fluxes between two graphene covered SiC plates. When graphene chemical potential increases from $\mu = 0$ eV to $\mu = 0.5$ eV, the first spectral heat flux peak becomes narrower and shifts from $\omega_1 = 4.5 \times 10^{13}$ rad/s to $\omega_1 = 1.3 \times 10^{14}$ rad/s, while the second one becomes broader, and also shifts from lower frequency of $\omega_2 = 1.8 \times 10^{14}$ rad/s to higher frequency of $\omega_2 = 2.7 \times 10^{14}$ rad/s. The shift of spectral heat flux peak to higher frequency with increasing graphene chemical potential is due to the graphene surface plasmon shift. This can be observed from the graphene optical properties as a function of frequency at different chemical potentials, which is shown in supporting material. One the other hand, all the spectral heat fluxes have a dip around $\omega_{dip} = 1.5 \times 10^{14}$ rad/s. The spectral heat flux between two bare SiC plates without graphene is also presented in Fig. 2. A very high narrow peak at $\omega_{SiC} = 1.78 \times 10^{14}$ rad/s can be clearly seen on the spectral heat flux between two bare SiC plates, which is known as the SiC SPhP coupling and has been well investigated [22, 25]. Through the comparison between the spectral heat fluxes for bare SiC plates and graphene covered ones, it can be concluded that the dip around $\omega_{dip} = 1.5 \times 10^{14}$ rad/s for the latter spectral heat flux is caused by the suppression of SiC SPhP modes from graphene surface plasmon. This is due to the fact that the graphene plasmon can suppress the heat flux modes with very large parallel wave vector supported by SiC SPhP modes, which is similar to the case of graphene covered SiO$_2$ plate and has been well illustrated [1]. The replacement of a single sharp spectral heat flux peak between two bare SiC plates with two broad ones after covering graphene sheets will also increase the total heat flux as shown in the following parts. Actually, similar to the phenomenon of the spectral heat flux peak splitting from one to two after covering graphene sheet on bare SiC plate, the splitting of local density of state (LDOS) at small distance from graphene covered SiC plate has already been observed from the dispersion relations of surface modes for graphene covered SiC [26].

The underlying mechanism for spectral heat flux peak splitting by adding a graphene layer and peak shifting with different graphene chemical potentials is explored through the contour plots of transmission coefficient presented in Fig. 3. The vacuum gap distance is still set at $d = 10$ nm, and Fig. 3(a-d) show the contour plot of transmission coefficient with graphene chemical potential of $\mu = 0$ eV, 0.1 eV, 0.3 eV, 0.5 eV, respectively. The bright area in the figure represents the transmission coefficient enhancement, which is equivalent to the heat flux improvement. As can be clearly observed from the figure, there are always two significantly enhanced transmission coefficient bands with four branches as graphene chemical potential varies. It is well known that two branches of SPhP coupling exist between two bare SiC plates around $\omega_{SiC} = 1.78 \times 10^{14}$ rad/s, which consists of symmetric low-frequency one and asymmetric high-frequency one [17, 22]. On the other hand, the SPP coupling between two bare graphene sheets have also been thoroughly investigated, which is a broadband enhancement also with two branches, and the resonance frequency strongly depends on graphene chemical potential [1, 14]. However, when SiC plates are covered with graphene sheets, as shown in Fig. 3, the splitting of transmission coefficient enhancement into four branches has not been observed before.

With increasing the graphene chemical potential from $\mu = 0$ eV to $\mu = 0.5$ eV, both enhancement bands shift to higher frequency. For the second band starting from around $\omega_2 = 2 \times 10^{14}$ rad/s, it will extend to around $\omega_2 = 5 \times 10^{14}$ rad/s at $\mu = 0.5$ eV. The enhancement band becomes broader and transmission coefficient peak shifts to higher frequency, which is consistent with the observation in Fig. 2. However, for the first band, it seems to have a saturation frequency at $\omega_{1sat} = 1.5 \times 10^{14}$ rad/s after $\mu = 0.3$ eV, which indicates that a further increase of chemical potential will not shift the enhancement band any more. After adding graphene sheet on SiC plate, the strong transmission coefficient enhancement at $\omega_{SiC} = 1.78\times10^{14}$ rad/s due to SiC SPhP coupling disappears, and the reason has been explained above as that SiC SPhP modes are suppressed by graphene surface plasmon. On the other hand, due to the suppression of SiC SPhP modes, the first transmission coefficient enhancement band caused by graphene SPP coupling has a saturation frequency around $\omega_{1sat} = 1.5\times10^{14}$ rad/s. The SPP coupling dispersion curves between two SiC-graphene-vacuum interfaces are also plotted in Fig. 3 by zeroing the denominator of transmission coefficient [17, 22, 27].

$$1 - r_{01}^p r_{02}^p e^{i2\gamma_0 d} = 0 \tag{5}$$

As presented in Fig. 3, there are always four SPP coupling dispersion curves as the graphene chemical potential changes. The four dispersion curves are perfectly matched to the four branches of transmission coefficient enhancement, which further confirms the effects of graphene SPP coupling with SiC substrates on improving the radiative heat transfer between them.

The total heat flux $q'' = \int_0^\infty q''_\omega d\omega$ is attained after integration of angular frequency and plotted as a function of chemical potential ($\mu_S = \mu_D$) for both graphene sheets in Fig. 4(a). Note that for numerical calculation of total heat flux, 500 data points have been used to integrate the spectral heat flux at the frequency range from $\omega = 1\times10^{13}$ rad/s to $\omega = 5\times10^{14}$ rad/s, and the convergence has been checked. When the chemical potential varies from 0 eV to 0.5 eV, the total heat flux increases at first to the maximum of $q''_{max} = 3.93$ MW/m$^2$ at $\mu_S = \mu_D = 0.15$ eV, and then decreases to the minimum $q''_{min} = 1.45$ MW/m$^2$ at $\mu_S = \mu_D = 0.5$ eV. Comparing with the total heat flux of $q''_{SiC} = 1.05$ MW/m$^2$ between two bare SiC plates at the gap distance of 10 nm, the total heat flux can be achieved as high as 3.75 times after covering graphene sheets on both SiC plates.

Let us now consider the proposed structure as a vacuum near-field thermal transistor with major functionalities of thermal modulation, amplification, and switching of heat flow between thermal source and drain. The dependence of near-field radiative heat flux on the graphene chemical potential has clearly shown the thermal modulation effect. To quantify the thermal amplification effect, the amplification factor $\Psi$ can be defined as

$$\Psi = \left| \partial q'' / \partial \mu_D \right| \quad (6)$$

which is essentially the slope of heat flux curve as a function of chemical potential. As plotted in Fig. 4(a), when the graphene chemical increases from 0 eV to 0.5 eV, the amplification factor increases at first to the maximum of $\Psi_{max}$ = 24.4 MW/(m$^2$ eV) at the chemical potential $\mu_S = \mu_D$ =0.06 eV, then decreases to $\Psi_{min}$ = 0 at $\mu_S = \mu_D = 0.15$ eV, where the total heat flux achieves its maximum value. Starting from $\mu_S = \mu_D = 0.15$ eV, the amplification increases at first from 0 to a second peak value of 10.5 MW/(m$^2$ eV) at $\mu_S = \mu_D = 0.23$ eV, then decreases to 3.9 MW/(m$^2$ eV) at $\mu_S = \mu_D = 0.5$ eV. Actually, we only need to vary the graphene chemical potential at the range from $\mu_S = \mu_D = 0$ eV to $\mu_S = \mu_D = 0.15$ eV to achieve the thermal amplification functionality with maximum amplification factor and or maximum total heat flux.

In order to illustrate the thermal switching effect, a switching factor is defined as

$$\Phi = 1 - q''_{min} / q''_{max} \quad (7)$$

The subscripts "max" and "min" refer to the maximum and minimum heat flux, which are defined as the switch on and off mode, respectively. From Fig. 4(a), the switching factor as large as $\Phi = 0.63$ can be gained at a vacuum gap distance $d = 10$ nm. At different vacuum gap distances, the maximum and minimum heat fluxes can always be obtained by varying graphene chemical potential. The maximum and minimum heat fluxes, as well as the switching factor as a function of vacuum gap distance are presented in Fig. 4(b). As the vacuum gap distance increases from $d = 10$ nm to $d = 1$ μm, the monotonic decrease of both maximum and minimum total heat fluxes is obvious because of the weaker SPP coupling at a larger gap distance, while a switching factor as large as $\Phi = 0.6$ can be sustained and almost unchanged at the vacuum gaps $d < 100$ nm. However, when $d > 100$ nm, the switching factor starts to drop down to $\Phi < 0.1$ at $d = 1$ μm. The decrease of switching factor at $d > 100$ nm is due to the weaker effect of graphene chemical potential to modulate the near-field radiative heat transfer as the SPP coupling becomes weaker at larger gap distances.

In order to obtain stronger thermal modulation, switching and amplification effects, we now consider different chemical potentials of the source and drain graphene sheets, i.e., asymmetric case with $\mu_S \neq \mu_D$. Figure 5(a) presents the contour plot of total near-field radiative heat flux between the thermal source and drain as a function of both the chemical potentials of graphene sheets changing from $\mu = 0$ eV to $\mu = 0.5$ eV at $d = 10$ nm. The heat flux is much higher when $\mu_S = \mu_D$ because the SPP coupling between them is strong due to the match of resonance frequencies. The contour plot in Fig. 5(a) is almost symmetric with respect to both the graphene chemical potentials, which is because the small temperature difference between the thermal source and drain hardly affects the dielectric function of graphene. The maximum heat flux $q''_{max} = 3.93$ MW/m$^2$ is still achieved at $\mu_S = \mu_D = 0.15$ eV, while the minimum heat flux $q''_{min} = 0.28$ MW/m$^2$ obtained at $\mu_S = 0.5$ eV and $\mu_D = 0.15$ eV for asymmetric case is much lower than that for symmetric case. Therefore, a larger switching factor of $\Phi = 0.93$ is attained. Obviously, the asymmetric structure provides more flexibility to modulate the heat flux between the thermal source and drain.

To characterize the amplification factor of thermal transistor with asymmetric structure, one of these two graphene chemical potentials needs to be fixed, and the heat flux change as a function of the other chemical potential can be explored. Figure 5(b) shows the amplification factor as a function of $\mu_D$ while $\mu_S$ is fixed. When $\mu_S$ is fixed at different values, there is always a dip close to 0 of $\Psi$ at $\mu_D = \mu_S$, which corresponds to the maximum total heat flux and $\Psi$ should be 0 due to the definition in Eq. (6). Interestingly, it is almost symmetric at the two sides of the dip position, which indicates that the total heat flux is nearly the same when $|\mu_S - \mu_D|$ equals. Compared to symmetric case, a larger maximum amplification factor of $\Psi_{max} = 59.2$ MW/(m$^2$ eV) is achieved at $\mu_S = 0.15$ eV and $\mu_D = 0.13$ eV.

In conclusion, we have theoretically demonstrated a near-field vacuum thermal transistor made of two graphene covered SiC plates, to modulate, amplify, and switch the heat flow by photon transport across a nanoscale vacuum gap. The graphene sheets function as gates by tuning its chemical potential with applied voltage biases. After adding graphene sheets on SiC plates, the near-field radiative heat transfer between them can be greatly enhanced due to graphene SPP coupling, while SiC SPhP coupling is suppressed. When the thermal source and drain temperatures are set as 400 K and 300K, respectively, the switching factor of $\Phi = 0.63$ and maximum amplification factor of $\Psi_{max} = 24.4$ MW/(m$^2$ eV) can be achieved at symmetric case of $\mu_S = \mu_D$, while $\Phi = 0.93$ and $\Psi_{max} = 59.2$ MW/(m$^2$ eV) is obtained for asymmetric case of $\mu_S \neq \mu_D$ at the vacuum gap distance $d = 10$ nm. This work will pave the way and open the thermal management system and future experiments.

YY and LW would like to thank the support from the ASU New Faculty Startup fund.

**References**

1. Li, B., L. Wang, and G. Casati, *Negative differential thermal resistance and thermal transistor.* Applied Physics Letters, 2006. **88**(14): p. 143501.
2. Chung Lo, W., L. Wang, and B. Li, *Thermal transistor: heat flux switching and modulating.* Journal of the Physical Society of Japan, 2008. **77**(5): p. 054402.
3. Ben-Abdallah, P. and S.-A. Biehs, *Near-field thermal transistor.* Physical Review Letters, 2014. **112**(4): p. 044301.
4. Otey, C.R., W.T. Lau, and S. Fan, *Thermal Rectification through Vacuum.* Physical Review Letters, 2010. **104**(15): p. 154301.
5. Iizuka, H. and S. Fan, *Rectification of evanescent heat transfer between dielectric-coated and uncoated silicon carbide plates.* Journal of Applied Physics, 2012. **112**(2): p. 024304.
6. Basu, S. and M. Francoeur, *Near-field radiative transfer based thermal rectification using doped silicon.* Applied Physics Letters, 2011. **98**(11): p. 113106.
7. Wang, L.P. and Z.M. Zhang, *Thermal Rectification Enabled by Near-field Radiative Heat Transfer Between Intrinsic Silicon and a Dissimilar Material.* Nanoscale and Microscale Thermophysical Engineering, 2013. **17**(4): pp. 337-348.
8. Yang, Y., S. Basu, and L. Wang, *Radiation-based near-field thermal rectification with phase transition materials.* Applied Physics Letters, 2013. **103**(16): p. 163101.
9. Falkovsky, L.A. and A.A. Varlamov, *Space-time dispersion of graphene conductivity.* The European Physical Journal B, 2007. **56**(4): pp. 281-284.
10. Falkovsky, L.A., *Optical properties of graphene.* Journal of Physics: Conference Series, 2008. **129**(1): p. 012004.
11. Falkovsky, L.A. and S.S. Pershoguba, *Optical far-infrared properties of a graphene monolayer and multilayer.* Physical Review B, 2007. **76**(15): p. 153410.
12. Yu, Y.-J., et al., *Tuning the Graphene Work Function by Electric Field Effect.* Nano Letters, 2009. **9**(10): pp. 3430-3434.
13. Svetovoy, V.B., P.J. van Zwol, and J. Chevrier, *Plasmon enhanced near-field radiative heat transfer for graphene covered dielectrics.* Physical Review B, 2012. **85**(15): p. 155418.
14. Ilic, O., et al., *Near-field thermal radiation transfer controlled by plasmons in graphene.* Physical Review B, 2012. **85**(15): p. 155422.
15. Lim, M., S.S. Lee, and B.J. Lee, *Near-field thermal radiation between graphene-covered doped silicon plates.* Optics Express, 2013. **21**(19): pp. 22173-22185.
16. Volokitin, A.I. and B.N.J. Persson, *Near-field radiative heat transfer between closely spaced graphene and amorphous $SiO_2$.* Physical Review B, 2011. **83**(24): p. 241407.
17. Liu, X. and Z. Zhang, *Graphene-assisted near-field radiative heat transfer between corrugated polar materials.* Applied Physics Letters, 2014. **104**(25): p. 251911.
18. Hu, L., et al., *Near-field thermal radiation between two closely spaced glass plates exceeding Planck's blackbody radiation law.* Applied Physics Letters, 2008. **92**(13): p. 133106.
19. Ganjeh, Y., et al., *A platform to parallelize planar surfaces and control their spatial separation with nanometer resolution.* Review of Scientific Instruments, 2012. **83**(10): p. 105101.
20. Shen, S., et al., *Nanoscale thermal radiation between two gold surfaces.* Applied Physics Letters, 2012. **100**(23): p. 233114.

21. Rytov, S.M., *Principles of Statistical Radiophysics*. 1987: Springer-Verlag.
22. Basu, S., Z.M. Zhang, and C.J. Fu, *Review of near-field thermal radiation and its application to energy conversion.* International Journal of Energy Research, 2009. **33**(13): pp. 1203-1232.
23. Fu, C. and Z. Zhang, *Nanoscale radiation heat transfer for silicon at different doping levels.* International Journal of Heat and Mass Transfer, 2006. **49**(9): pp. 1703-1718.
24. Alaee, R., et al., *A perfect absorber made a graphene micro-ribbon metamaterial.* Optics Express, 2012. **20**(27): pp. 28017-28024.
25. Greffet, J.-J., et al., *Coherent emission of light by thermal sources.* Nature, 2002. **416**(6876): p. 61-64.
26. Messina, R., et al., *Tuning the electromagnetic local density of states in graphene-covered systems via strong coupling with graphene plasmons.* Physical Review B, 2013. **87**(8): p. 085421.
27. Basu, S. and L. Wang, *Near-field radiative heat transfer between doped silicon nanowire arrays.* Applied Physics Letters, 2013. **102**(5): p. 053101.

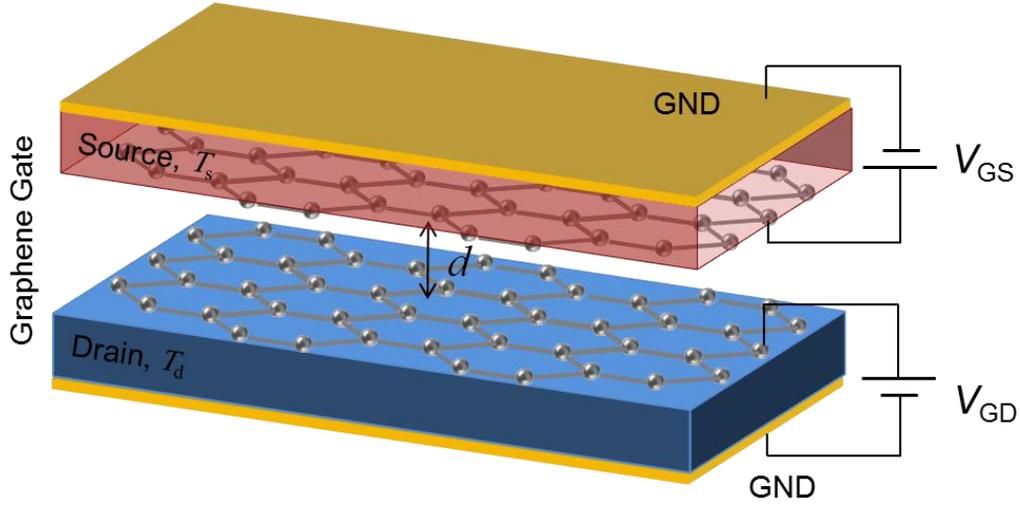

FIG. 1 Schematic of near-field vacuum thermal transistor consisting of two graphene covered SiC plates. The two SiC plates and graphene sheets represent the thermal source, drain and gate, respectively. The vacuum gap distance is denoted as $d$, which is set as $d = 10$ nm in this study, and the source and drain temperatures are set as $T_S = 400$ K and $T_D = 300$ K respectively. Two metal plates are also added to serve as ground electrodes, for which $V_{GS}$ and $V_{GD}$ are applied to respectively tune the source and drain graphene chemical potentials.

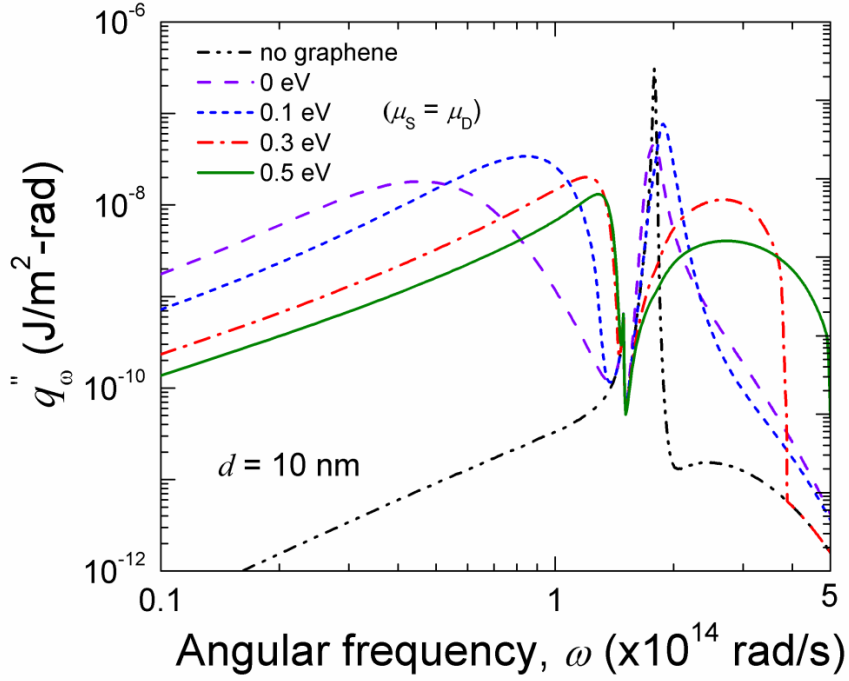

FIG. 2 Spectral heat fluxes between source and drain when both the graphene sheets take the same chemical potential. The spectral heat flux between two bare SiC plates without graphene is also plotted.

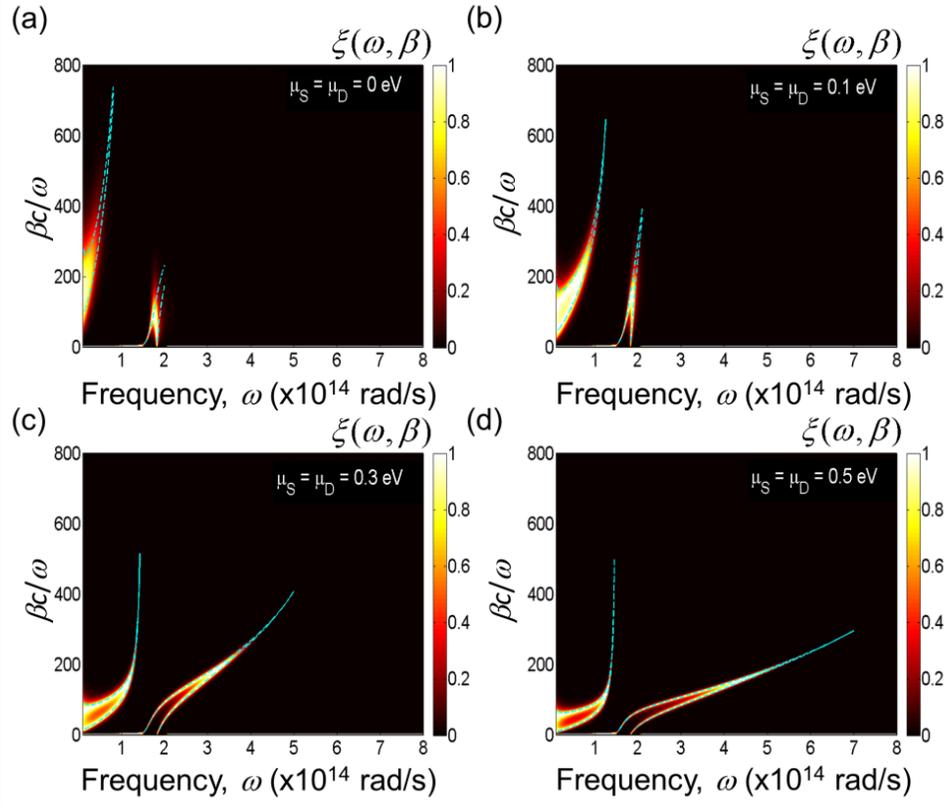

FIG. 3 Contour plots of the transmission coefficient between source and drain with different chemical potentials applied to both the graphene sheets. The SPP coupling dispersion curves are also plotted to match the enhancement.

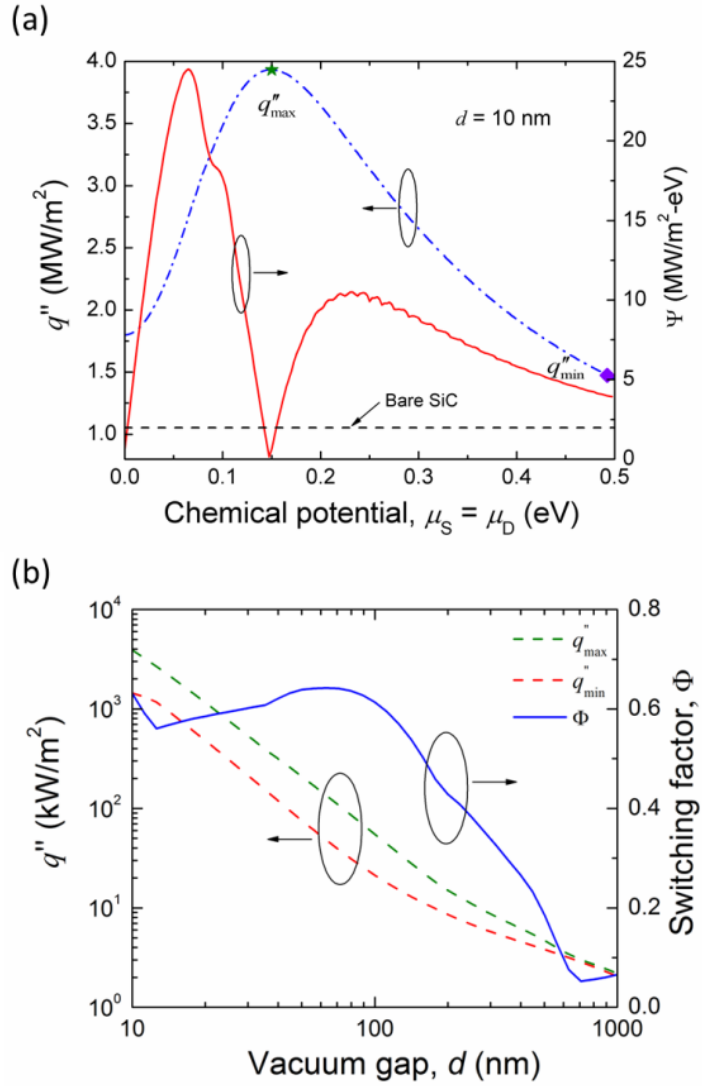

FIG. 4. (a) Total heat flux between source and drain (blue dash line) and thermal amplification factor (red solid line) versus chemical potentials applied to both graphene sheets. (b) Thermal switching factor as a function of vacuum gap distance.

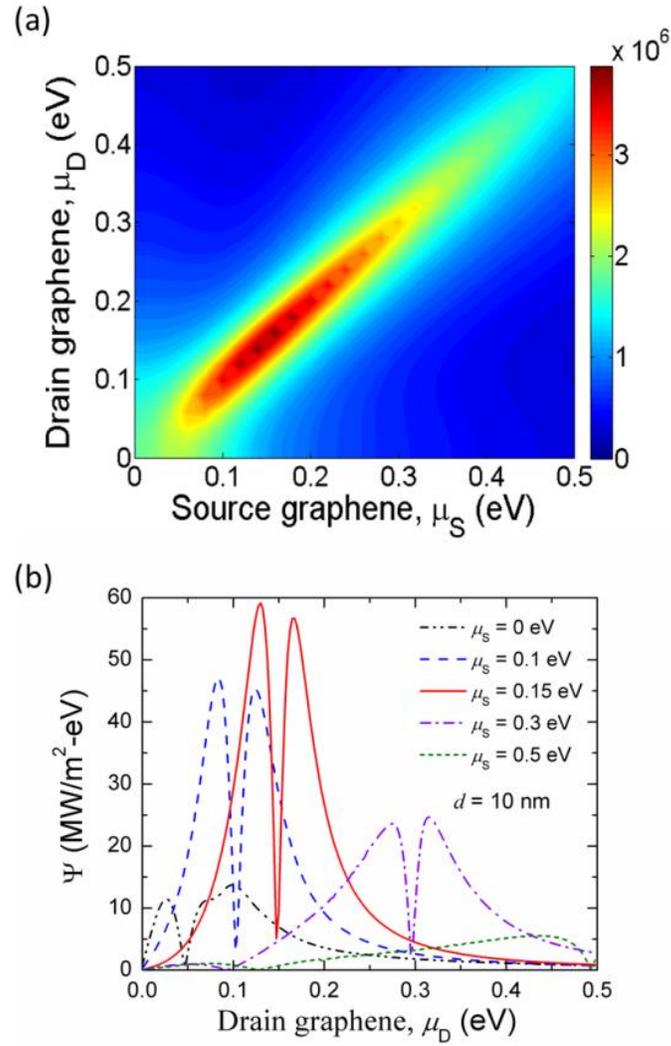

FIG. 5 (a) Contour plot of heat flux between source and drain for the asymmetric case and as a function of the different chemical potentials applied to the two graphene sheets. (b) Thermal amplification factor vs the chemical potential applied for the graphene sheet covering on the drain when that for the thermal source is fixed.

17